# Impact of Optic Nerve Tortuosity, Globe Proptosis, and Size on Retinal Ganglion Cell Thickness Across General, Glaucoma, and Myopic Populations

Charis Y.N. Chiang[1,2,3], Xiaofei Wang[4], Stuart K. Gardiner[5], Martin Buist[2], Michaël J.A. Girard[1,3,6,7,8]

1. Department of Ophthalmology, Emory University School of Medicine, Atlanta, Georgia USA
2. Department of Biomedical Engineering, National University of Singapore, Singapore
3. Singapore Eye Research Institute, Singapore National Eye Centre, Singapore
4. Key Laboratory for Biomechanics and Mechanobiology of Ministry of Education, Beijing Advanced Innovation Center for Biomedical Engineering, School of Biological Science and Medical Engineering, Beihang University, Beijing, China
5. Devers Eye Institute, Legacy Health, Portland, Oregon, United States
6. Department of Biomedical Engineering, Georgia Institute of Technology/Emory University, Atlanta, GA, USA
7. Emory Empathetic AI for Health Institute, Emory University, Atlanta, GA, USA
8. Duke-NUS Graduate Medical School, Singapore

**Word count:** 4067 (manuscript text)

**Tables:** 4
+11 (supplementary material)

**Figures:** 6
+4 (supplementary material)

**Commercial relationship:** MJAG is the co-founder of the start-up company Abyss Processing Pte Ltd.

**Ethics statement:** This study was covered by the generic ethical approval for UK Biobank studies from the National Research Ethics Service Committee North West–Haydock (approval letter dated 29 June 2021, Ref 21/NW/0157), and all study procedures adhered to the tenets of the Declaration of Helsinki. Written informed consent was obtained from each subject.

**Support:** Girard acknowledges support from Emory Eye Center (Start-up funds, MJAG), BrightFocus Foundation (G2021010S), TARGET (MOH-OFLCG21jun-0003) and Challenge Grant from Research to Prevent Blindness, Inc.

**Corresponding Author:** Dr Michaël J.A. Girard
Ophthalmic Engineering & Innovation Laboratory (OEIL), Emory Eye Center, Emory School of Medicine, Emory Clinic Building B, 1365B Clifton Road, NE, Atlanta GA 303
mgirard@ophthalmic.engineering
https://www.ophthalmic.engineering

# ABSTRACT

**Purpose**: To investigate the impact of optic nerve tortuosity (ONT), and the interaction of globe proptosis and size on retinal ganglion cell (RGC) thickness, using Retinal Nerve Fiber Layer (RNFL) thickness, across general, glaucoma, and myopic populations.

**Methods**: This study analyzed 17940 eyes from the UKBiobank cohort (ID 76442), including 72 glaucomatous and 2475 myopic eyes. Artificial intelligence models were developed to derive RNFL thickness corrected for ocular magnification from 3D optical coherence tomography scans and orbit features from 3D magnetic resonance images including ONT, globe proptosis, axial length, and a novel feature: the interzygomatic line-to-posterior pole (ILPP) distance – a composite marker of globe proptosis and size. Generalized estimating equation (GEE) models evaluated associations between orbital and retinal features.

**Results**: RNFL thickness was positively correlated with ONT and ILPP distance ($r = 0.065$, $p < 0.001$, and $r = 0.206$, $p < 0.001$ respectively) in the general population. The same was true for glaucoma ($r = 0.040$, $p = 0.74$, and $r = 0.224$, $p = 0.059$), and for myopia ($r = 0.069$, $p < 0.001$, and $r = 0.100$, $p < 0.001$). GEE models revealed that straighter optic nerves and shorter ILPP distance were predictive of thinner RNFL in all populations.

**Conclusions**: Straighter optic nerves and decreased ILPP distance could cause RNFL thinning, possibly due to greater traction forces. ILPP distance emerged as a potential biomarker of axonal health. These findings underscore the importance of orbit structures in RGC axonal health and warrant further research into orbit biomechanics.

# INTRODUCTION

The optic nerve is essential for transmitting visual information from retinal ganglion cells (RGC) to the brain, and multiple studies suggest that mechanical forces acting on it may significantly impact RGC axonal health [1]. These studies, using techniques such as Optical Coherence Tomography (OCT), Finite Element modelling, and Magnetic Resonance Imaging (MRI), have identified the optic nerve head as particularly susceptible to deformation caused by optic nerve traction, a pulling force on the optic nerve induced by eye movements [2–10]. This traction is likely influenced by the structural characteristics of the orbit, which may, in turn, potentially compromise RGC axonal integrity, affecting both the nerve and the retinal nerve fiber layer (RNFL) [7]. However, despite growing interest in this area, the interplay between the orbital structures and RGC axonal health remains poorly understood.

Understanding the factors influencing traction forces is particularly relevant to glaucoma, a leading cause of irreversible blindness, characterized by progressive loss of RGC axons and visual field loss [11,12]. Previous studies have highlighted two key orbital features associated with glaucoma: optic nerve geometry and globe size and/or position. First, Wang et al. found that glaucoma patients had less tortuous optic nerves across primary, abduction and adduction gazes compared to controls [8]. Straighter nerves could increase RGC axonal stress during eye movements. The same study noted increased globe proptosis at primary gaze, which may elevate tension at the optic nerve head and exacerbate glaucomatous damage [8]. Demer et al. reported abnormal globe retraction due to optic nerve tethering during adduction in glaucoma subjects, particularly in Asian eyes [7,10]. This may induce stress at the globe-optic nerve junction, aligning with Wang et al.'s findings of increased proptosis in Chinese glaucoma subjects [7,10]. Similarly, thyroid eye disease, a risk factor of glaucoma, has been associated with altered orbital mechanics and increased optic nerve strain in eyes with proptosed globes, reinforcing the potential impact of orbital influences on RGC axonal health [6,13].

Additionally, optic nerve traction may partially contribute to myopia development. Wang et al. found that estimated optic nerve tortuosity (ONT) before myopic onset, was significantly lower in highly myopic eyes than measured ONT in controls suggesting that higher traction forces occur in eyes predisposed to myopia [9]. Chuangsuwanich et al. reported exaggerated intraocular pressure-induced strain in myopic eyes and positive correlation between adduction-induced strain at the and intraocular pressure-induced strain [14]. Given this and that optic nerve head connective tissues are weakened in myopic eyes [15,16], it is plausible that adduction-induced strain at the optic nerve head is elevated in myopia. These increased forces at the optic nerve head could contribute to RGC axonal loss in myopic, providing a potential explanation for myopia being a risk factor for glaucoma [16].

Despite these findings, the full extent of how orbital structures influence axonal health remains underexplored, particularly in glaucoma and myopia. Notably, no studies to date have systematically examined the correlation between orbital structures and RGC axonal health. The UK Biobank initiative offers a unique opportunity to investigate these associations on a large scale with extensive population-wide imaging data, including brain MRIs (capturing the globe and orbit) and macular OCTs. Additionally, advances in artificial intelligence (AI) allow for the large-scale analysis of these complex relationships, Together, these facilitate the study of previously underexplored biomechanical factors through a population-wide study.

In this study, we aimed to elucidate the relationship between axonal health, surrogated by macular RNFL thickness, and specific orbital features. RNFL thickness is a non-invasive marker for axonal health, as thinner RNFLs indicate greater axonal degeneration [17,18]. Additionally, we aimed to investigate the complex interactions among orbital parameters, and introduced a novel orbital feature that may provide a stronger correlation with RNFL thickness and offer potential for predicting its changes.

# METHODS

## Dataset Curation

This population study used demographic, diagnostic and imaging data from the UKBiobank cohort (ID 76442). The UK Biobank study is a large-scale, ongoing population study following 500,000 adult UK residents from 2010 onward. Disease information for subjects was encoded using the International Classification of Diseases (ICD-9 and ICD-10) codes. A subset of participants underwent macular OCT scans and structural brain MRIs.

The 3D spectral domain OCT scans were centered on the fovea, covering a 6mm x 6mm area using a raster scan protocol (3D OCT-1000 Mark II, Topcon, Japan). Each OCT volume was stored in FDS format and consisted of 128 B-scans at 47.2µm resolution, 512 A-scans per B-scan at 11.7µm resolution and 650 pixels per A-scan at an axial resolution of 3.5µm.

The T1-weighted structural brain MRI scans were acquired using a Siemens Skyra 3T system running VD13A SP4 (as of October 2015) with a 32-channel RF head coil. Scans covered the full head in 208 sagittal slices, each 256 x 256 pixels at a 1mm isotropic voxel resolution, over a 5-minute duration.

We curated a dataset to include subjects with both an OCT scan (in at least one eye) and an MRI scan. Sex, disease diagnoses, and age were recorded. The glaucoma subpopulation was identified based on the ICD-10 code H40.1, corresponding to primary open-angle glaucoma (POAG). Throughout this text, all references to the glaucoma population will here on refer to subjects diagnosed with POAG. Given the distinct pathophysiological mechanisms underlying different glaucoma types, we conducted separate analyses for other glaucoma types with these results presented in Supplementary Note 1. We defined eyes with axial myopia, where axial length was greater than or equal to 25 mm, as belonging to the myopic subpopulation.

This study was covered by the generic ethical approval for UK Biobank studies from the National Research Ethics Service Committee North West–Haydock (approval letter dated 29

June 2021, Ref 21/NW/0157), and all study procedures adhered to the tenets of the Declaration of Helsinki. Written informed consent was obtained from each subject.

## Measurement of Retinal Nerve Fiber Layer Thickness

To assess RNFL thickness, we developed a robust AI algorithm for segmenting retinal tissue structures from OCT scans before measuring RNFL thickness around the fovea. To this end, a total of 120 OCT B-scan images were randomly selected from 41 scan volumes across 27 subjects. Scans were image-compensated to improve contrast and tissue visibility using the methods described in our previous study [19]. The compensated OCT images were then manually segmented (Figure 1a) to identify the following tissue groups: (1) RNFL, (2) ganglion cell layer, (3) inner plexiform layer, (4) inner nuclear layer, (5) outer plexiform layer, (6) outer nuclear layer, (7) external limiting membrane, (8-11) four photoreceptor layers, (12) retinal pigment epithelium, and (13) choroid, with the background assigned a value of zero. In most images, the posterior choroid boundary was indistinct, and the sclera was not visible due to limited depth penetration; only visible tissues were segmented.

The manually segmented dataset was split into 85% for training and validation and 15% for testing. A cross-entropy loss function was used. As per the methods of our previous study [19], extensive data augmentation was applied to the training set, including horizontal flipping, random rotation and translation, additive Gaussian noise, and random saturations. A UNET++ model was then trained to automatically segment the tissue layers in the OCT images. Five-fold cross validation was performed, and the mean Dice coefficient metric was used to evaluate model performance.

RNFL thickness was measured as the minimum distance between the anterior and posterior boundaries (Figure 1b), multiplying pixel counts by pixel resolution for true measurements. The foveal center was identified using a modified gradient descent algorithm applied to a thickness map of the four inner layers. Average global RNFL thickness was calculated within a 1.5 mm radius of the fovea, excluding the central 0.5 mm, and for the superior, nasal, inferior, and temporal quadrants (Figure 1c).

## Extraction of Orbit Features

To analyze the features in the orbit, we developed custom AI algorithms for automatic segmentation of key orbit structures from MRI scans and the subsequent extraction of relevant features from the segmented scans.

**Segmentation of MRI Scans.** MRI scans were cropped around each globe center to capture the entire globe, zygomatic bone, and visible optic nerve. The images were then resampled to 0.2 mm isotropic resolution using a Lanczos3 filter for precise segmentation, and left-eye volumes were horizontally flipped to match the orientation of right-eye volumes.

From a randomly selected subset of 15 subjects (30 orbits), seven orbital structures were manually segmented in 3D using the Amira software: (1) optic nerve, (2) globe, (3) zygomatic bone, (4) lens, (5) scleral-corneal shell, (6) rectus muscles, and (7) orbital fat (Figure 2a). The 6910 MRI slices were divided into 80% for training and validation and 20% for testing, ensuring slices from the same subject remained in the same set. Data augmentation was applied to the training data, and then a UNET model, trained with four-fold cross-validation, was used to automatically segment the entire dataset.

After segmenting the left and right eyes from each scan, the cropped-out eye orbits were reassembled to form a complete scan. The central points through the globe and lens in coronal slices were identified and fit to a central orbital axis for each eye using linear regression. This method was applicable even in myopic eyes as, to the best of our knowledge, none exhibited staphyloma. The central axial plane was then established using three points: the first two being the anterior corneal points of the left and right central orbital axes, and the third point being the midpoint between the left and right posterior poles as depicted in Figure 2b.

**Axial Length.** Axial length (AL) was estimated from MRI, using the distance between the cornea and posterior globe along the central orbital axis: which joined the lens and globe center [8,9]. As such, after identifying the axial plane, AL was measured as the distance between the anterior corneal point and the posterior pole along the central orbital axis.

**Globe Proptosis.** Globe proptosis for each eye was calculated by measuring the distance between the anterior corneal point and the interzygomatic line. The interzygomatic line connects the most anterior points of the left and right zygomatic bone in the central axial plane.

**Interzygomatic Line-to-Posterior Pole Distance.** We introduced a new feature: the Interzygomatic Line-to-Posterior Pole Distance (ILPP) distance, defined as axial length minus globe proptosis. This proposed composite marker was designed to capture the interaction between globe size and proptosis. We hypothesized that ILPP distance would more closely relate to the amount of optic nerve traction experienced and provide a more nuanced understanding of orbital structure and biomechanics, and their relationship to axonal health.

**Optic Nerve Tortuosity.** The segmented optic nerve was skeletonized by computing the mean x and y coordinates of the optic nerve in each coronal slice to locate its center. The line was fitted to a cubic spline to reduce inter-slice noise. A 20 mm intraorbital segment was extracted beginning at the junction of the sclera and nerve, and ONT was calculated as 20 mm divided by the straight-line distance between the segment endpoints (Figure 3).

## RNFL Thickness Corrected for Ocular Magnification

To correct for ocular magnification, we employed the popular Littman formula, as modified by Bennett, to adjust the RNFL thickness values. The retinal feature true size ($t$) was calculated by multiplying the imaging system's magnification factor ($p$), the ocular magnification factor ($q$), and the measured size ($s$); i.e. $t = p \times q \times s$ [20–22]. For the Topcon system, $p$ is 3.382 [22]. The ocular magnification factor was calculated using the formula $q = 0.01306 \times (axial\ length - 1.82)$ [20–22].

## Quality Control

OCT and MRI pairs with corrupted files or poor-quality scans were excluded. In the glaucoma and myopia datasets, scans hindering retinal or orbital structure visualization were manually removed following manual checks by 3D visualization. For the general population, outliers were removed using thresholding based on both Tukey's definition and biological plausibility [23]. Eyes deviating from primary gaze were also excluded using the methods described in Supplementary Note 2. Extracted features were manually reviewed in 3D for all glaucoma and myopic subsets and selected general population subjects, with corrections applied as needed.

## Statistical Analysis

Statistical analysis was conducted to evaluate the relationships between orbital features and RNFL thickness. Bivariate correlation analysis was used to assess the association between RNFL thickness and orbital features.

Generalized estimating equations (GEE) were employed to understand the combined effects of ONT and globe position and size on RNFL thickness, adjusting for age (mean of age at time of OCT and at time of MRI scan). GEE accounts for within-subject correlations in eye measurements. To minimize empirical and model-based variance estimates, an exchangeable working correlation matrix was employed. Data was normalized before modeling. Features with a tolerance value (reciprocal of variance inflation factor) below 0.6, indicating strong multicollinearity, were excluded from the GEE. P-values below 0.05 were deemed statistically significant.

# RESULTS

The UK Biobank contains 27996 pairs of OCT and MRI scans from 14617 subjects, of which 491 pairs from 250 subjects had a glaucoma diagnosis (of any type) based on ICD codes. Exclusion factors were corruption of files during data extraction, poor quality data, processing errors, failure of manual checks, deviation from primary gaze, and thresholding. The final tally was 17940 pairs of scans with 72 pairs from POAG subjects, and 2475 pairs from myopic subjects. The demographic data is reported in Table 1.

The segmentation networks for both OCT and MRI achieved good performance, with Dice coefficients of 0.947 ± 0.003 and 0.954 ± 0.001, respectively. These trained networks were then used to segment all the OCT and MRI scans before feature extraction algorithms extracted the orbital and retinal features, as shown in Table 2 and Figure 4.

**General Population.** In the general population, correlation analysis revealed significant positive linear relationships between ONT and corrected global RNFL thickness ($r = 0.065$, $p < 0.001$) and between ILPP distance and corrected global RNFL thickness ($r = 0.206$, $p < 0.001$) (Table 3; Figure 5). The latter was significantly stronger according to Steiger's test ($p < 0.001$) [24]. The GEE model revealed ONT, ILPP distance, and age as significant predictors of corrected global RNFL thickness in the general population ($\beta = 0.022$, $p < 0.001$, $\beta = 0.205$, $p < 0.001$, $\beta = -0.100$, $p < 0.001$, respectively) (Table 4). These results are depicted in Figure 6.

**Glaucoma Sub-population.** In the glaucoma subpopulation, ONT was positively but statistically insignificantly correlated with corrected global RNFL thickness ($r = 0.040$, $p = 0.74$) as shown in Table 3 and Figure 5. ILPP distance also showed a positive correlation with corrected global RNFL thickness ($r = 0.224$, $p = 0.059$). The GEE model results are depicted in Figure 6. The coefficients of ONT and ILPP distance as predictors of corrected global RNFL thickness were positive but statistically insignificant. (Table 4),

**Myopic Subpopulation.** In the myopic subpopulation, ONT was significantly positively correlated with corrected global RNFL thickness ($r = 0.069$, $p < 0.001$) as shown in Table 3 and Figure 5. ILPP distance also showed a significant positive correlation with corrected global RNFL thickness ($r = 0.100$, $p < 0.001$). The GEE model results are depicted in Figure 6. The coefficient of ILPP distance as a predictor of corrected global RNFL thickness was statistically significant ($\beta = 0.108$, $p < 0.001$), though the coefficient of ONT was not ($\beta = 0.029$, $p = 0.10$) (Table 4),

As ILPP distance and globe proptosis exhibited multicollinearity (tolerance $< 0.4$), these features were not included together in any regression model (Tables 4 – 6). Models including globe proptosis can be found in the Supplementary Material. All other pairs of independent variables had tolerance values greater than 0.6, indicating no high correlation between them. The prediction models for uncorrected global RNFL and for RNFL divided into each of the four quadrants can also be found in the Supplementary Material.

# DISCUSSION

In this study, we conducted the first population-wide investigation into orbital features and their relationship with axonal health, using corrected global RNFL thickness as a surrogate marker. Our analysis employed an AI-based medical image analysis framework with feature-based extraction to examine orbital anatomy. A novel composite metric, the ILPP distance, was introduced to account for globe proptosis and size. The results demonstrated significant positive correlations between corrected global RNFL thickness and both ONT and ILPP distance. Furthermore, these two orbital features emerged as significant predictors of corrected global RNFL thickness in GEE regression models, highlighting their potential relevance in understanding axonal health. These findings suggest that there are critical links between the structure and possibly even biomechanics of the orbit and the retina.

Our study revealed significant correlations between less tortuous optic nerves and thinner corrected global RNFL in the general population, similar trends were observed in both disease populations. This finding aligns with biomechanical principles: a straighter optic nerve has less redundancy during eye adduction, becoming taut earlier [25]. Once this redundancy is exhausted, the optic nerve, its sheath, and other orbital tissues likely experience greater traction forces, requiring them to stretch to accommodate eye movements [25,26]. These increased forces may contribute to axonal damage, which could lead to RNFL thinning, aligning with previous studies suggesting that mechanical tension and strain on the optic nerve influence axonal health and RNFL thickness [7–9]. Such findings highlight the role of nerve geometry in maintaining optic nerve function and its potential effect on individual susceptibility to neurodegeneration, supporting ONT as a biomarker to predict RGC axonal damage.

Moreover, we found significant positive correlations between ILPP distance and corrected global RNFL thicknesses. In the GEE models, the coefficient for ILPP distance as a predictor of corrected global RNFL thickness was also significantly positive. These findings suggest that a smaller ILPP distance, indicative of a smaller posterior segment, could induce greater traction forces on the RGC axons. We propose two potential explanations for this observation. First, a smaller ILPP distance might cause the globe-nerve junction to be more anteriorly placed. This could reduce optic nerve redundancy, impair eye movements and exacerbate RNFL thinning. This could be particularly so for glaucomatous subjects, especially during adduction, where this effect may lead to tethering as observed by Demer et al., causing globe retraction and increasing traction forces on the optic nerve, especially at the optic nerve head [10]. Second, a reduced ILPP distance is correlated with increased globe proptosis, which may hinder eye movements and elevate traction forces experienced at the posterior retina. This aligns with the findings of Fisher et al., who reported increased optic nerve strain in subjects with proptosed globes [6]. It also complements the work of Wang et al., who found that both glaucoma and myopic patients exhibited more globe proptosis than healthy subjects [8,9], Together, these studies suggest that globe proptosis may contribute to axonal damage by increasing the mechanical strain on the optic nerve and retina.

Correction of RNFL thickness using axial length had minimal impact on the GEE model results but strengthened the weakly positive correlation between RNFL thickness and axial length. This finding contrasts with prior studies reporting weak negative correlation pre-correction and weak positive correlation post correction [21,22,27]. Methodological differences, participant demographics, or specific OCT devices employed may contribute to this discrepancy. Though we do note the contention that RNFL thinning by myopic axial elongation poses to this result [27,28], the positive correlation may reflect interactions between ILPP distance and RNFL thickness. Specifically, with the same globe proptosis, an eye with longer axial length may have a more posteriorly placed globe-nerve junction, potentially allowing more optic nerve redundancy, allowing reduced traction forces and releasing stress, thus resulting in less RNFL thinning. Further investigation is needed to better understand the complex interplay between ocular anatomy, axial length, and RNFL measurements.

In this study, several limitations warrant further discussion. First, the quality of the UK Biobank imaging data was compromised by artifacts such as shadows, blurring, and misalignment, which introduced noise. Despite excluding scans via manual checks and outlier removal, some localized errors in 3D image volumes might have gone undetected. Future studies could incorporate advanced quality control algorithms to eliminate noise, improve the overall data quality and potentially strengthen the current observed trends.

Another limitation of the UK Biobank data collection is that many subjects had their eyes closed during MRI scanning, and subjects were not specifically instructed to fixate their eyes at primary gaze. We developed an algorithm to sieve out and exclude such scans not at primary gaze but note that there could still be tiny deviations within plus or minus five degrees. This could affect the measurements taken from the MRI scans, particularly ONT, adding noise to the data. However, with a large cohort, this noise could be negligible for such an exploratory study identifying preliminary trends. While the data might not be perfectly clean, it provides a strong foundation for further investigation and a scan protocol designed specifically to examine eyes in primary gaze could possibly strengthen the observed trends.

Third, the automatic segmentation algorithms were trained on a limited sample relative to the large dataset, potentially propagating errors into the feature extraction process. In particular, though 6910 image slices were used to train the MRI segmentation algorithm, these came from only 15 subjects. Additionally, multiple custom feature extraction algorithms were applied to a large volume of data. While we aimed for consistency, the sheer volume of data made it impractical to manually check the entire general population dataset for errors. Future analyses with more robust segmentation algorithms trained on larger datasets, along with advanced automated algorithms to detect segmentation errors and other data inconsistencies could strengthen the observed trends and provide even more reliable insights into the relationship between orbit features and RNFL thickness.

Fourth, the statistical methods used in this study, including correlation and linear GEE regression, were limited to identifying linear relationships. For instance, although axial length appeared to have the strongest linear correlation with RNFL thickness a separate test using mutual information, as detailed in Supplementary Note 3, suggested that ILPP distance, globe proptosis, and ONT exhibited higher dependence with RNFL thickness than axial length. This implies that the relationships between RNFL thickness and these features could be non-linear in nature. While this is a limitation, our analysis effectively captured the overall trends, which are sufficient for the scope of this investigation. Future research could incorporate more advanced AI techniques to better explore these complex interactions in the orbit.

Fifth, there is a lack of information on whether glaucoma was unilateral or bilateral, and thus unilateral cases could potentially be included in the dataset. However, this limitation is mitigated by the fact that glaucoma is predominantly a bilateral disease, with studies reporting up to 90% of glaucoma patients affected bilaterally and contralateral conversion rates reaching 41.7% within two years for unilateral cases [29–33].

Finally, diagnosis information was limited to recorded ICD codes which significantly limited the disease-specific analyses, particularly within the glaucoma population. Therefore, detailed information on disease severity and subtype was unavailable. As such, glaucoma severity, which may significantly influence RNFL thickness, was unaccounted for in our analysis. Disease classification in the UK Biobank included self-reported data, which may limit the reliability of glaucoma diagnoses. Furthermore, over half of the participants with a glaucoma diagnosis were assigned the ICD-10 code H40.9 (unspecified glaucoma), making it likely that several POAG cases were not captured in our analysis. Myopia was also severely underreported, necessitating the definition based on axial length. Nonetheless, the association between thinner RNFLs and both straighter optic nerves and decreased ILPP distance in glaucoma and myopic patients remains. Normal-tension and high-tension glaucoma could also differently influence the relationship between orbital structures and retinal neurodegeneration. Future studies could be crafted to incorporate subtype-specific analysis and include disease severity which may reveal how different biomechanical forces, such as variations in ILPP distance or axial length, affect RNFL thickness in a disease-specific manner.

In conclusion, this study highlights the importance of expanding neuro-ophthalmology research to include orbital structures and biomechanics as critical factors in axonal health. Our findings suggest that ONT, globe size and proptosis play significant roles in axonal health and could provide new avenues for understanding neurodegeneration, particularly in diseases such as glaucoma and myopia. Future research could delve further into understanding the orbit features and its impact on axonal health. These could include measuring ONT in different gaze positions and exploring the biomechanical effects on the optic nerve head and exploring its links with other features such as head size, brain volume, genetics and other factors to further advance our understanding of these mechanisms

# FIGURES

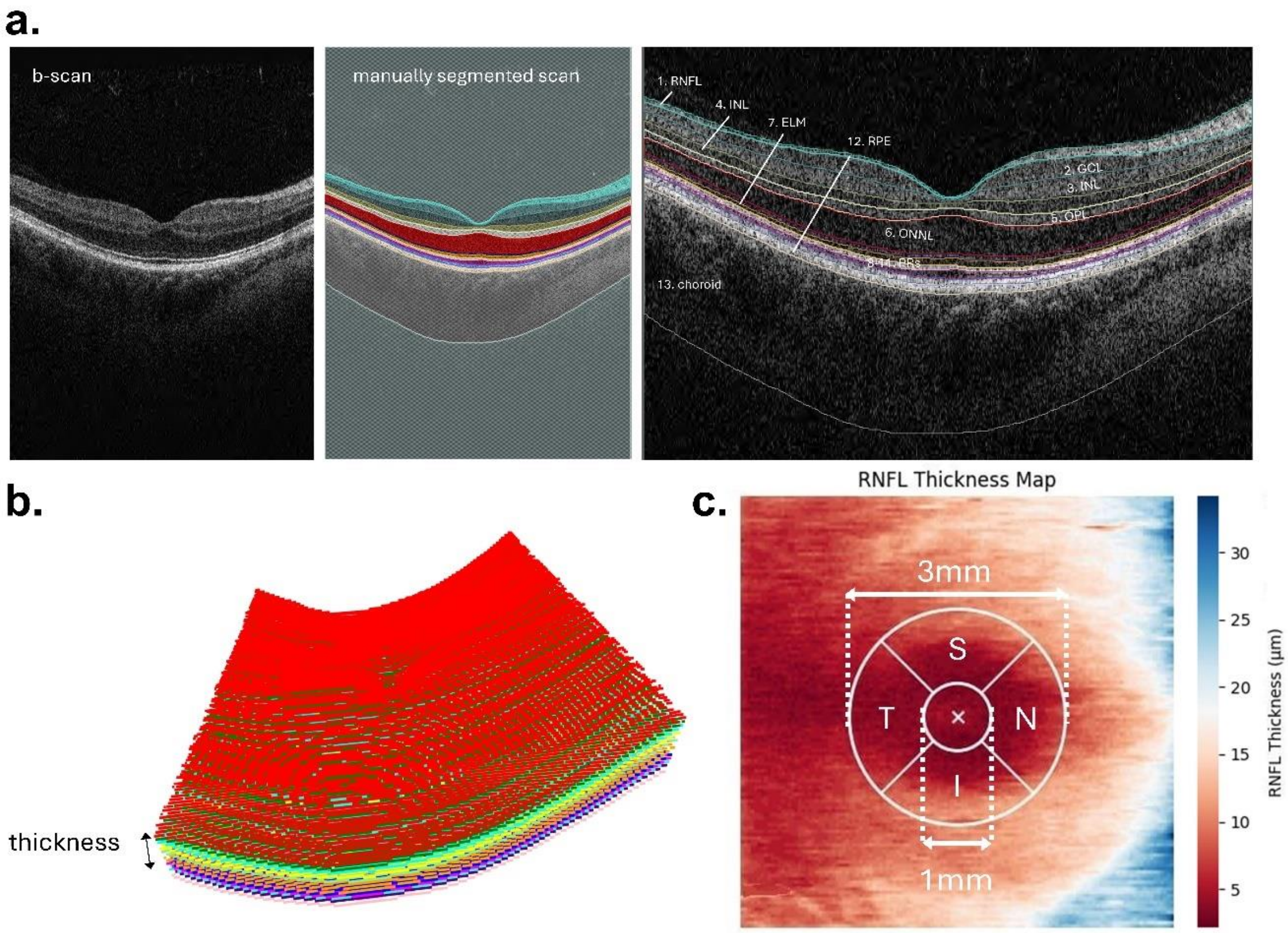

**Figure 1 a.** Thirteen tissue layers were manually segmented in B-scan slices from the optical coherence tomography volumes: (1) retinal nerve fiber layer (RNFL), (2) ganglion cell layer, (3) inner plexiform layer, (4) inner nuclear layer, (5) outer plexiform layer, (6) outer nuclear layer, (7) external limiting membrane, (8-11) four photoreceptor layers, (12) retinal pigment epithelium, and (13) choroid **b.** Segmented b-scans were pieced together to form the volume RNFL thickness was measured at every point of the scan area. **c.** The average thicknesses of the superior (S), nasal (N), inferior (I), and temporal (T) quadrants, and globally around the fovea were taken.

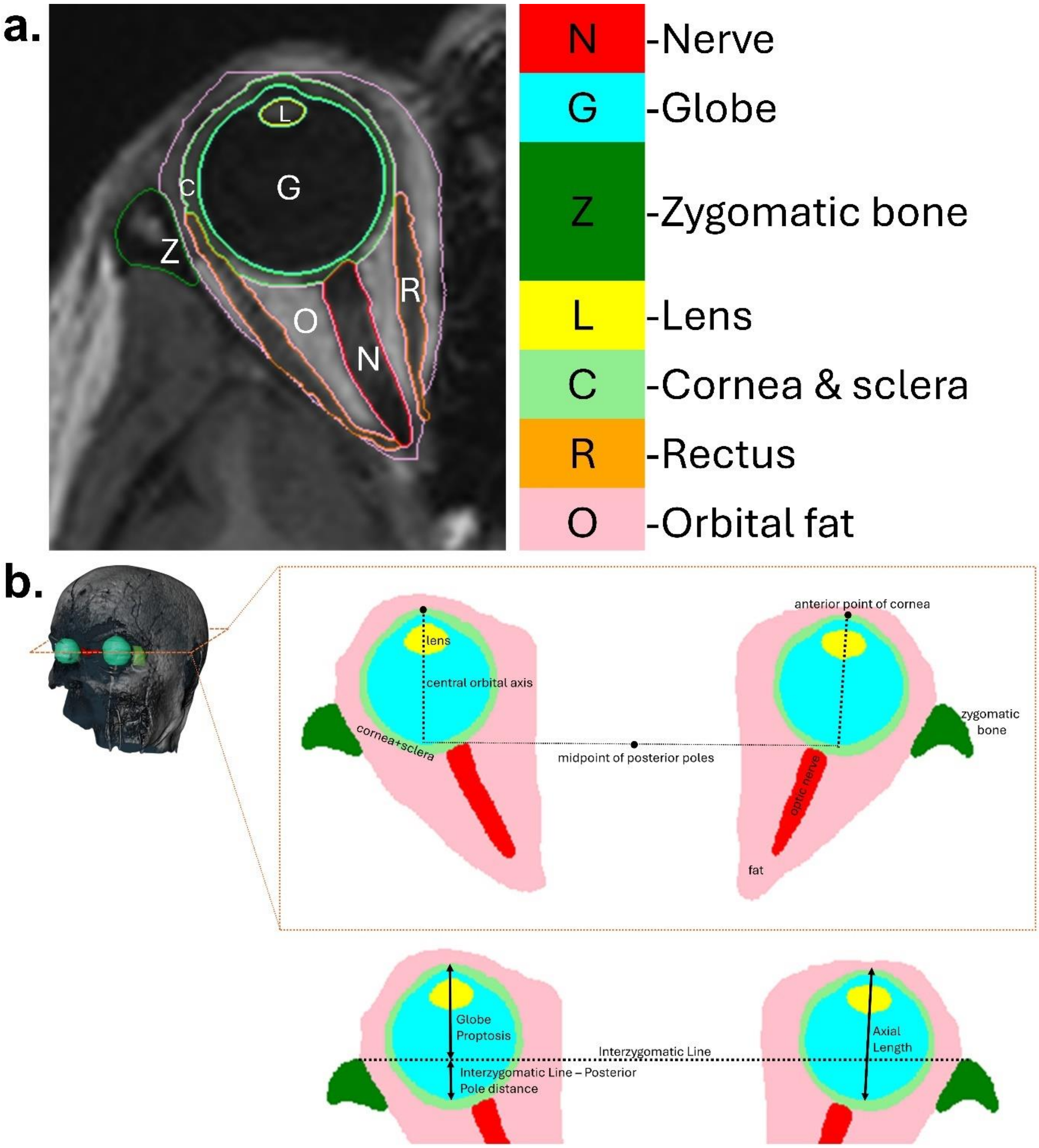

**Figure 2a.** Seven orbital structures were manually segmented in cropped-out orbit portions of 3D T1 brain magnetic resonance images **b.** The central axial plane was found using the anterior points of each cornea and midpoint of the posterior poles. From there, axial length was measured as the distance between the anterior cornea and posterior pole. Globe proptosis was measured from the anterior cornea to the interzygomatic line. Interzygomatic line-to-posterior pole distance was calculated as axial length minus globe proptosis.

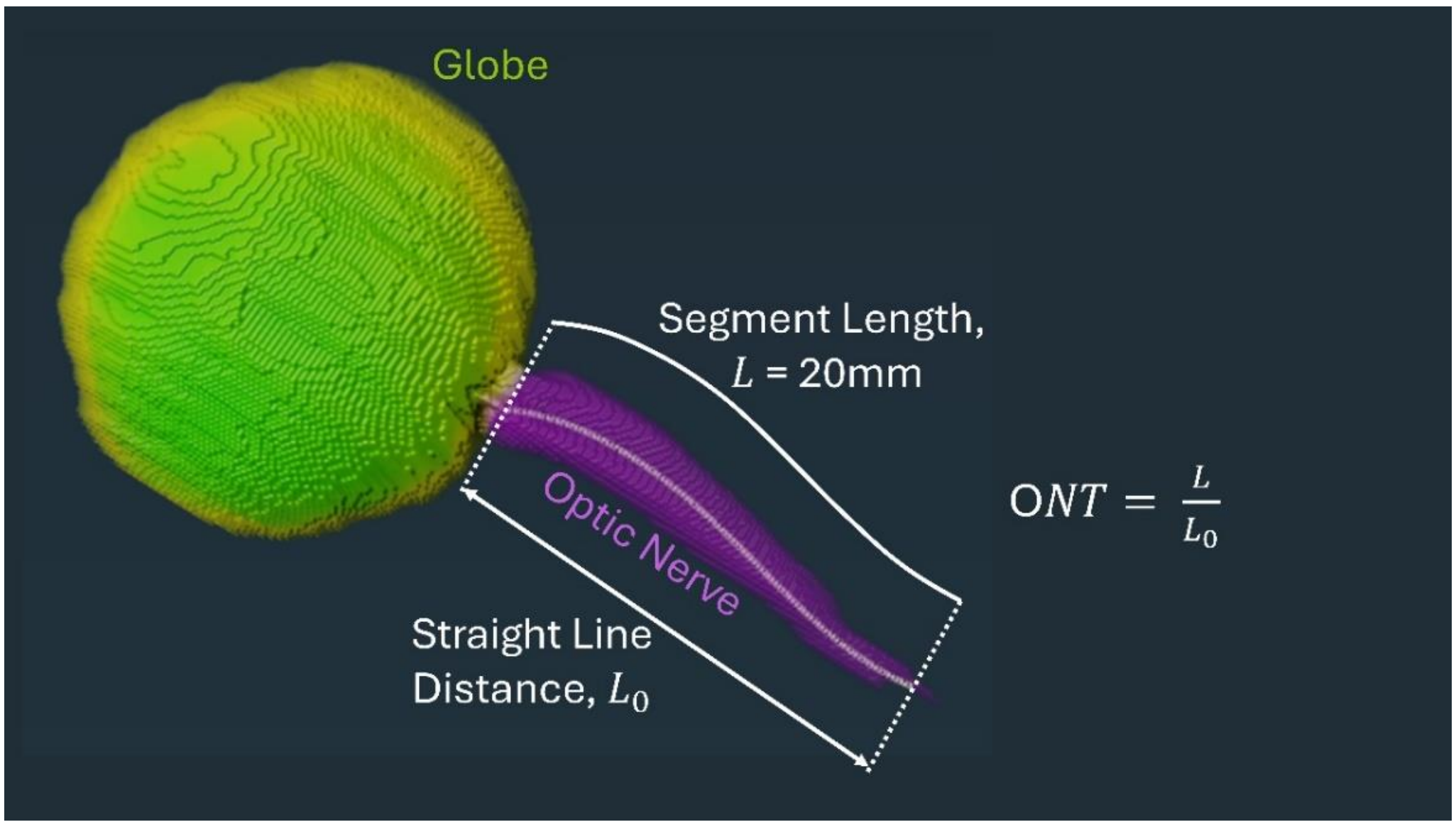

**Figure 3.** Optic nerve tortuosity (ONT) was taken as 20mm divided by the straight-line distance from end-to-end of a 20mm anterior segment of the skeletonized optic nerve.

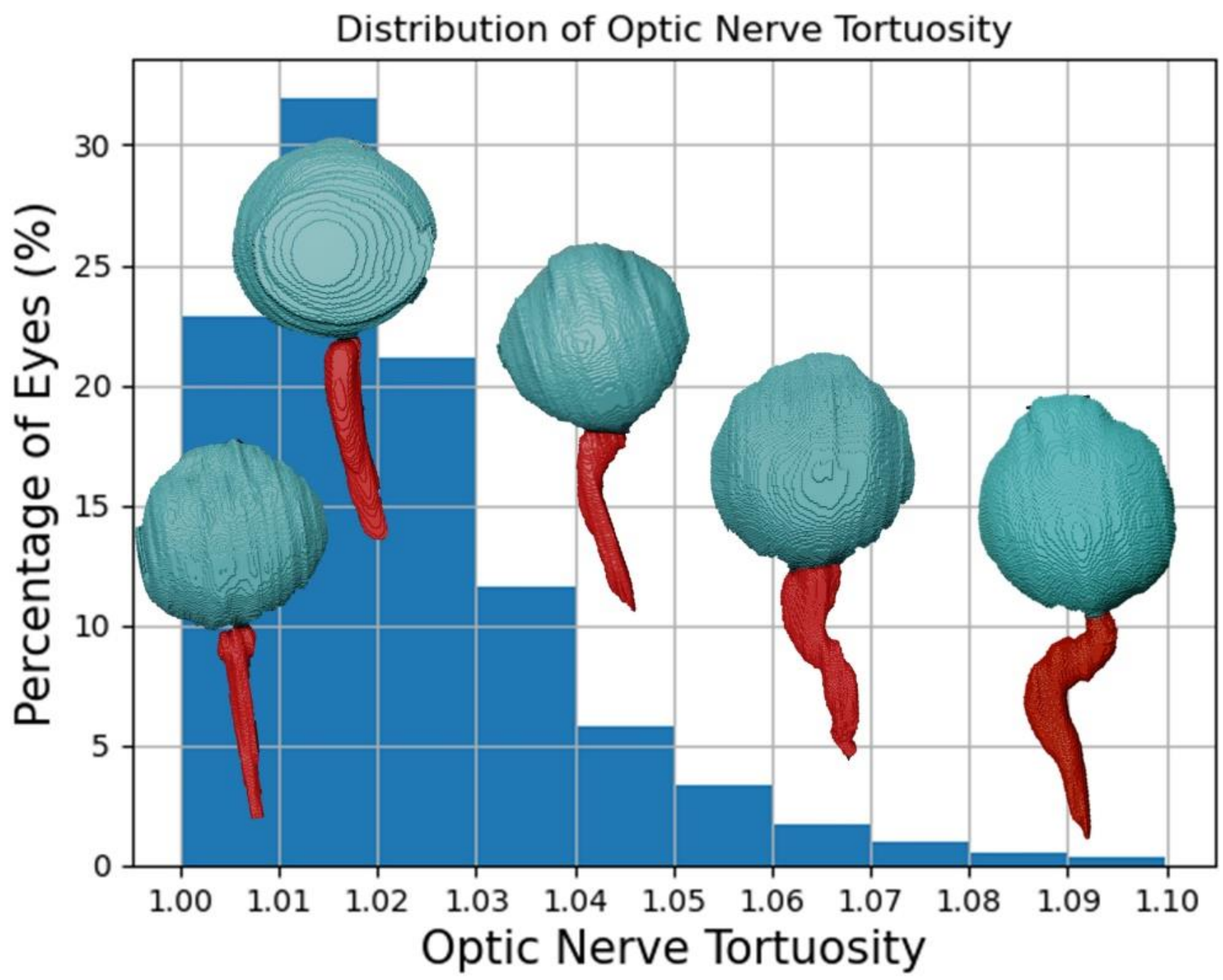

**Figure 4.** Distribution of optic nerve tortuosity and depiction of optic nerve geometry for different tortuosity values

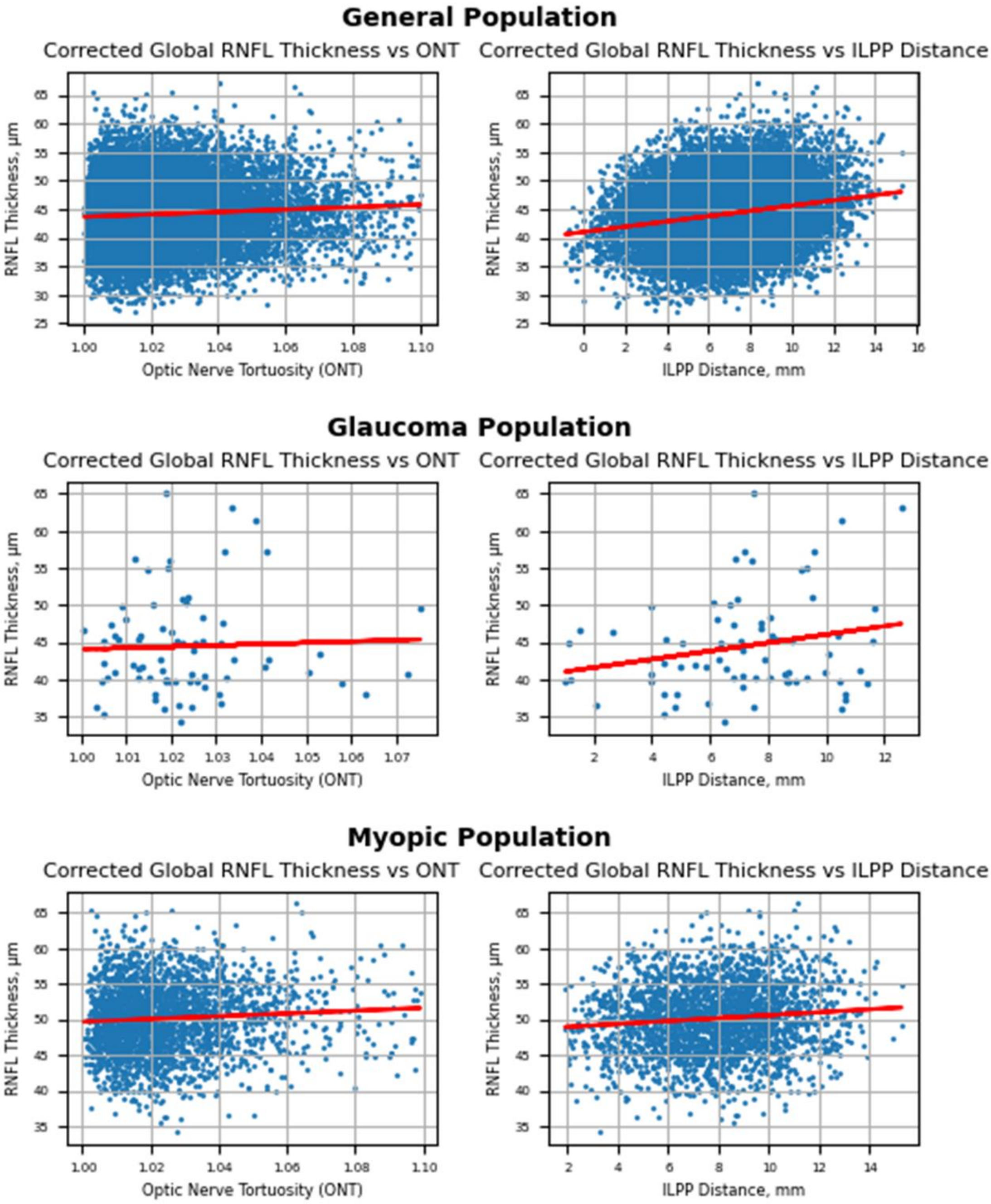

**Figure 5.** Corrected Global RNFL Thickness plotted against ONT and ILPP Distance in the general population and glaucoma and myopic subpopulation. RNFL, retinal nerve fiber layer; ONT, optic nerve tortuosity; ILPP, interzygomatic line-to-posterior pole.

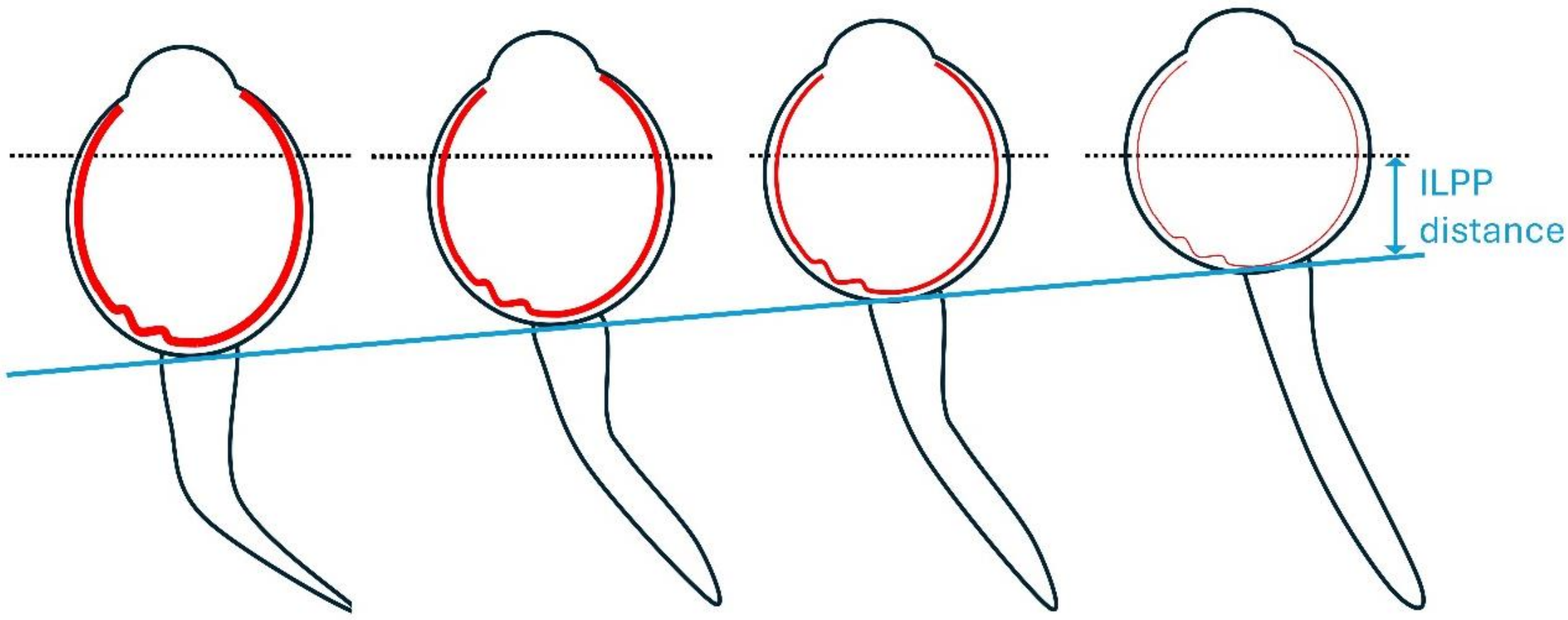

**Figure 6.** Retinal Nerve Fiber Layer Thinning is associated with decreased interzygomatic line-to-posterior pole (ILPP) Distance and reduced optic nerve tortuosity in the general population and both disease subpopulations.

**TABLES**

**Table 1.** Demographics of subjects from the General Population and Glaucoma Subset

| | **General Population** | **Glaucoma Subpopulation** | **Myopic Subpopulation** |
|---|---|---|---|
| Number of participants | 11508 | 44 | 1800 |
| Gender | 5749 Male, 5759 Female | 25 Male, 19 Female | 1170 Male, 630 Female |
| Number of scan pairs | 17940 | 72 | 2475 |
| Age at OCT scan, years | 58.042 ± 7.842 | 63.694 ± 5.631 | 57.751 ± 7.725 |
| Age at MRI scan, years | 64.546 ± 7.759 | 70.097 ± 5.977 | 64.090 ± 7.791 |

**Table 2**. Summary of Retinal and Orbital Features

| **Feature** | **General Population** | **Glaucoma Subpopulation** | **Myopic Subpopulation** |
|---|---|---|---|
| Optic Nerve Tortuosity | 1.022 ± 0.016 | 1.023 ± 0.015 | 1.024 ± 0.017 |
| ILPP Distance, mm | 6.798 ± 2.353 | 7.145 ± 2.715 | 7.820 ± 2.367 |
| Globe Proptosis, mm | 16.824 ± 2.367 | 17.170 ± 2.521 | 17.886 ± 2.339 |
| Axial Length, mm | 23.619 ± 1.208 | 24.315 ± 1.510 | 25.706 ± 0.701 |
| Age, years | 61.619 ± 7.671 | 66.896 ± 5.614 | 60.921 ± 7.629 |
| Corrected Global RNFL thickness, μm | 44.153 ± 5.258 | 44.480 ± 6.758 | 50.151 ± 4.840 |

**Table 3**. Correlation of Corrected Global RNFL Thickness with various Orbital Features

| | **General Population** | | **Glaucoma Subpopulation** | | **Myopic Subpopulation** | |
|---|---|---|---|---|---|---|
| **Orbital Feature** | **Pearson r** | **p-value** | **Pearson r** | **p-value** | **Pearson r** | **p-value** |
| Optic Nerve Tortuosity | 0.065 | $3 \times 10^{-18}$ | 0.040 | 0.74 | 0.069 | $6 \times 10^{-4}$ |
| ILPP Distance | 0.206 | $9 \times 10^{-171}$ | 0.224 | 0.059 | 0.100 | $6 \times 10^{-7}$ |
| Globe Proptosis | 0.140 | $1 \times 10^{-79}$ | 0.149 | 0.213 | 0.013 | 0.53 |
| Axial Length | 0.675 | $7 \times 10^{-2377}$ | 0.650 | $6 \times 10^{-10}$ | 0.370 | $6 \times 10^{-81}$ |
| Age | -0.090 | $7 \times 10^{-34}$ | 0.006 | 0.96 | -0.069 | $6 \times 10^{-4}$ |

**Table 4**. Linear GEE Regression Results to Predict Corrected Global RNFL Thickness

| | **Coef.** | **Std Err** | **z** | **P>\|z\|** | **95% CI** | |
|---|---|---|---|---|---|---|
| | **General Population** | | | | | |
| Optic Nerve Tortuosity | 0.022 | 0.006 | 3.803 | $1 \times 10^{-4}$ | 0.011 | 0.034 |
| ILPP Distance | 0.205 | 0.009 | 23.464 | $1 \times 10^{-121}$ | 0.187 | 0.221 |
| Age | -0.100 | 0.010 | -9.577 | $1 \times 10^{-21}$ | -0.120 | -0.08 |
| | **Glaucoma Subpopulation** | | | | | |
| Optic Nerve Tortuosity | 0.056 | 0.082 | 0.687 | 0.49 | -0.104 | 0.217 |
| ILPP Distance | 0.308 | 0.171 | 1.805 | 0.07 | -0.026 | 0.642 |
| Age | -0.021 | 0.136 | -0.153 | 0.88 | -0.288 | 0.246 |
| | **Myopic Subpopulation** | | | | | |
| Optic Nerve Tortuosity | 0.029 | 0.017 | 1.666 | 0.10 | -0.005 | 0.063 |
| ILPP Distance | 0.108 | 0.023 | 4.721 | $6 \times 10^{-6}$ | 0.063 | 0.153 |
| Age | -0.066 | 0.025 | -2.575 | 0.01 | -0.116 | -0.016 |

# ACKNOWLEDGEMENTS

Acknowledgement is made to **(1)** This research has been conducted using data from UK Biobank, a major biomedical database; **(2)** Emory Eye Center [Start-up funds, MJAG]; **(3)** the donors of the National Glaucoma Research, a program of the BrightFocus Foundation, for support of this research (G2021010S [MJAG]); **(4)** NMRC-LCG grant 'TAckling & Reducing Glaucoma Blindness with Emerging Technologies (TARGET)', award ID: t [MJAG]; and **(5)** Support from a Challenge Grant from Research to Prevent Blindness, Inc.

**SUPPLEMENTARY MATERIAL 1**

**Supplementary Table 1**. Correlation of Global Retinal Nerve Fiber Layer (RNFL) Thickness with various Orbital Features

| | **General Population** | | **Glaucoma Subpopulation *** | | **Myopic Subpopulation** | |
|---|---|---|---|---|---|---|
| **Orbital Feature** | **Pearson r** | **p-value** | **Pearson r** | **p-value** | **Pearson r** | **p-value** |
| Optic Nerve Tortuosity | 0.049 | $6 \times 10^{-11}$ | 0.003 | 0.98 | 0.054 | 0.007 |
| ILPP Distance | 0.122 | $3 \times 10^{-60}$ | 0.046 | 0.70 | 0.050 | 0.01 |
| Globe Proptosis | 0.019 | 0.010 | 0.098 | 0.41 | -0.068 | 0.19 |
| Axial Length | 0.275 | $1 \times 10^{-309}$ | 0.246 | 0.04 | 0.071 | $4 \times 10^{-4}$ |
| Age | -0.102 | $9 \times 10^{-43}$ | 0.134 | 0.26 | -0.074 | $2 \times 10^{-4}$ |

**Glaucoma Subpopulation comprises subjects with primary open-angle glaucoma*

**Supplementary Table 2**. Correlation of Retinal Nerve Fiber Layer (RNFL) Thicknesses in 4 Quadrants with Orbit Features in the General Population

| Retinal Feature | Orbital Feature | Pearson r | p-value |
|---|---|---|---|
| Superior RNFL Thickness | Optic Nerve Tortuosity | 0.047 | $4 \times 10^{-10}$ |
| | ILPP Distance | 0.100 | $2 \times 10^{-41}$ |
| | Globe Proptosis | 0.005 | 0.46 |
| | Axial Length | 0.205 | $9 \times 10^{-170}$ |
| | Age | -0.079 | $3 \times 10^{-26}$ |
| Corrected Superior RNFL Thickness | Optic Nerve Tortuosity | 0.062 | $5 \times 10^{-17}$ |
| | ILPP Distance | 0.180 | $3 \times 10^{-131}$ |
| | Globe Proptosis | 0.111 | $2 \times 10^{-50}$ |
| | Axial Length | 0.569 | $2 \times 10^{-1530}$ |
| | Age | -0.076 | $3 \times 10^{-24}$ |
| Inferior RNFL Thickness | Optic Nerve Tortuosity | 0.037 | $6 \times 10^{-7}$ |
| | ILPP Distance | 0.088 | $2 \times 10^{-32}$ |
| | Globe Proptosis | 0.015 | 0.05 |
| | Axial Length | 0.200 | $8 \times 10^{-163}$ |
| | Age | -0.109 | $2 \times 10^{-48}$ |
| Corrected Inferior RNFL Thickness | Optic Nerve Tortuosity | 0.054 | $5 \times 10^{-13}$ |
| | ILPP Distance | 0.166 | $1 \times 10^{-110}$ |
| | Globe Proptosis | 0.113 | $4 \times 10^{-52}$ |
| | Axial Length | 0.544 | $3 \times 10^{-1379}$ |
| | Age | -0.102 | $4 \times 10^{-43}$ |
| Temporal RNFL Thickness | Optic Nerve Tortuosity | 0.035 | $3 \times 10^{-6}$ |
| | ILPP Distance | 0.051 | $1 \times 10^{-11}$ |
| | Globe Proptosis | 0.031 | $4 \times 10^{-5}$ |
| | Axial Length | 0.158 | $2 \times 10^{-100}$ |
| | Age | -0.043 | $9 \times 10^{-9}$ |
| Corrected Temporal RNFL Thickness | Optic Nerve Tortuosity | 0.058 | $5 \times 10^{-15}$ |
| | ILPP Distance | 0.173 | $4 \times 10^{-120}$ |
| | Globe Proptosis | 0.173 | $8 \times 10^{-121}$ |
| | Axial Length | 0.675 | $7 \times 10^{-2377}$ |
| | Age | -0.047 | $4 \times 10^{-10}$ |
| Nasal RNFL Thickness | Optic Nerve Tortuosity | 0.040 | $1 \times 10^{-7}$ |
| | ILPP Distance | 0.145 | $2 \times 10^{-84}$ |
| | Globe Proptosis | 0.023 | 0.002 |
| | Axial Length | 0.327 | $5 \times 10^{-448}$ |
| | Age | -0.082 | $6 \times 10^{-28}$ |
| Corrected Nasal RNFL Thickness | Optic Nerve Tortuosity | 0.055 | $1 \times 10^{-13}$ |
| | ILPP Distance | 0.216 | $3 \times 10^{-182}$ |
| | Globe Proptosis | 0.126 | $5 \times 10^{-64}$ |
| | Axial Length | 0.660 | $6 \times 10^{-2236}$ |
| | Age | -0.075 | $9 \times 10^{-24}$ |

**Supplementary Table 3**. Linear Generalized Estimating Equation Regression Models to Predict Global Retinal Nerve Fiber Layer (RNFL) Thickness

| | Coef. | Std Err | z | P>\|z\| | 95% CI | |
|---|---|---|---|---|---|---|
| | **General Population** | | | | | |
| Optic Nerve Tortuosity | 0.024 | 0.006 | 3.825 | $1 \times 10^{-4}$ | 0.011 | 0.036 |
| ILPP Distance | 0.109 | 0.009 | 12.502 | $7 \times 10^{-36}$ | 0.091 | 0.125 |
| Age | -0.111 | 0.010 | -10.896 | $1 \times 10^{-27}$ | -0.131 | -0.091 |
| | **Glaucoma Subpopulation *** | | | | | |
| Optic Nerve Tortuosity | 0.008 | 0.062 | 0.134 | 0.89 | -0.113 | 0.130 |
| ILPP Distance | 0.108 | 0.137 | 0.789 | 0.43 | -0.160 | 0.376 |
| Age | 0.093 | 0.113 | 0.820 | 0.41 | -0.129 | 0.315 |
| | **Myopic Subpopulation** | | | | | |
| Optic Nerve Tortuosity | 0.026 | 0.018 | 1.482 | 0.138 | -0.008 | 0.060 |
| ILPP Distance | 0.051 | 0.023 | 2.233 | 0.03 | 0.006 | 0.096 |
| Age | -0.070 | 0.025 | -2.752 | 0.006 | -0.120 | -0.020 |

**Glaucoma Subpopulation comprises subjects with primary open-angle glaucoma*

**Supplementary Table 4**. Linear Generalized Estimating Equation Regression Models to Predict Global Retinal Nerve Fiber Layer (RNFL) Thickness using Globe Proptosis and Axial Length in the General Population

| | Coef. | Std Err | z | P>\|z\| | 95% CI | |
|---|---|---|---|---|---|---|
| | **Prediction of Global RNFL Thickness** | | | | | |
| Optic Nerve Tortuosity | 0.023 | 0.006 | 3.835 | $1 \times 10^{-4}$ | 0.011 | 0.035 |
| Globe Proptosis | -0.046 | 0.009 | -5.298 | $1 \times 10^{-7}$ | -0.063 | -0.029 |
| Axial Length | 0.268 | 0.009 | 30.673 | $1 \times 10^{-206}$ | 0.250 | 0.285 |
| Age | -0.106 | 0.01 | -10.781 | $4 \times 10^{-27}$ | -0.126 | -0.087 |
| | **Prediction of Corrected Global RNFL Thickness** | | | | | |
| Optic Nerve Tortuosity | 0.023 | 0.005 | 4.735 | $2 \times 10^{-6}$ | 0.013 | 0.032 |
| Globe Proptosis | -0.035 | 0.007 | -5.066 | $4 \times 10^{-7}$ | -0.048 | -0.021 |
| Axial Length | 0.672 | 0.007 | 95.327 | 0 | 0.658 | 0.686 |
| Age | -0.086 | 0.008 | -11.154 | $7 \times 10^{-29}$ | -0.102 | -0.071 |

**Supplementary Table 5** Linear Generalized Estimating Equation Regression Models to Predict Global Retinal Nerve Fiber Layer (RNFL) Thickness using Globe Proptosis and Axial Length in the Primary Open-Angle Glaucoma Subpopulation

| | Coef. | Std Err | z | P>\|z\| | 95% CI | |
|---|---|---|---|---|---|---|
| | **Prediction of Global RNFL Thickness** | | | | | |
| Optic Nerve Tortuosity | -0.009 | 0.056 | -0.157 | 0.875 | -0.119 | 0.101 |
| Globe Proptosis | -0.025 | 0.125 | -0.201 | 0.841 | -0.271 | 0.220 |
| Axial Length | 0.265 | 0.112 | 2.373 | 0.018 | 0.046 | 0.483 |
| Age | 0.145 | 0.116 | 1.241 | 0.215 | -0.084 | 0.373 |
| | **Prediction of Corrected Global RNFL Thickness** | | | | | |
| Optic Nerve Tortuosity | 0.008 | 0.055 | 0.141 | 0.888 | -0.099 | 0.115 |
| Globe Proptosis | -0.027 | 0.124 | -0.219 | 0.826 | -0.271 | 0.217 |
| Axial Length | 0.753 | 0.118 | 6.398 | 0.000 | 0.523 | 0.984 |
| Age | 0.131 | 0.117 | 1.117 | 0.264 | -0.099 | 0.361 |

**Supplementary Table 6**. Linear Generalized Estimating Equation Regression Models to Predict Global Retinal Nerve Fiber Layer (RNFL) Thickness using Globe Proptosis and Axial Length in the Myopic Subpopulation

| | Coef. | Std Err | z | P>\|z\| | 95% CI | |
|---|---|---|---|---|---|---|
| | **Prediction of Global RNFL Thickness** | | | | | |
| Optic Nerve Tortuosity | 0.024 | 0.018 | 1.345 | 0.18 | -0.011 | 0.058 |
| Globe Proptosis | -0.033 | 0.023 | -1.401 | 0.16 | -0.078 | 0.013 |
| Axial Length | 0.055 | 0.02 | 2.724 | 0.006 | 0.015 | 0.094 |
| Age | -0.069 | 0.025 | -2.715 | 0.007 | -0.119 | -0.019 |

| | Prediction of Corrected Global RNFL Thickness | | | | | |
|---|---|---|---|---|---|---|
| Optic Nerve Tortuosity | 0.011 | 0.017 | 0.63 | 0.53 | -0.022 | 0.043 |
| Globe Proptosis | -0.034 | 0.022 | -1.539 | 0.12 | -0.077 | 0.009 |
| Axial Length | 0.295 | 0.02 | 14.715 | $5 \times 10^{-49}$ | 0.255 | 0.334 |
| Age | -0.059 | 0.024 | -2.446 | 0.01 | -0.106 | -0.012 |

**Supplementary Table 7**. Linear Generalized Estimating Equation Regression Models to Predict Retinal Nerve Fiber Layer (RNFL) Thickness in 4 Quadrants in the General Population

| | Coef. | Std Err | z | P>\|z\| | 95% CI | |
|---|---|---|---|---|---|---|
| | **Prediction of Superior RNFL Thickness** | | | | | |
| Optic Nerve Tortuosity | 0.025 | 0.007 | 3.555 | $4 \times 10^{-4}$ | 0.011 | 0.037 |
| ILPP Distance | 0.090 | 0.009 | 10.241 | $1 \times 10^{-24}$ | 0.073 | 0.107 |
| Age | -0.088 | 0.01 | -8.767 | $2 \times 10^{-18}$ | -0.109 | -0.069 |
| | **Prediction of Corrected Superior RNFL Thickness** | | | | | |
| Optic Nerve Tortuosity | 0.023 | 0.007 | 3.383 | 0.001 | 0.009 | 0.035 |
| ILPP Distance | 0.179 | 0.009 | 19.929 | $2 \times 10^{-88}$ | 0.161 | 0.196 |
| Age | -0.086 | 0.01 | -8.224 | $2 \times 10^{-16}$ | -0.106 | -0.065 |
| | **Prediction of Inferior RNFL Thickness** | | | | | |
| Optic Nerve Tortuosity | 0.021 | 0.007 | 3.019 | 0.003 | 0.007 | 0.034 |
| ILPP Distance | 0.079 | 0.009 | 8.723 | 0.078 | 0.061 | 0.096 |
| Age | -0.124 | 0.01 | -11.962 | -0.124 | -0.144 | -0.103 |
| | **Prediction of Corrected Inferior RNFL Thickness** | | | | | |
| Optic Nerve Tortuosity | 0.024 | 0.007 | 3.538 | $4 \times 10^{-4}$ | 0.01 | 0.037 |
| ILPP Distance | 0.160 | 0.009 | 18.124 | $2 \times 10^{-73}$ | 0.143 | 0.178 |
| Age | -0.116 | 0.01 | -11.246 | $2 \times 10^{-29}$ | -0.136 | -0.096 |
| | **Prediction of Temporal RNFL Thickness** | | | | | |
| Optic Nerve Tortuosity | 0.028 | 0.007 | 4.141 | $3 \times 10^{-5}$ | 0.014 | 0.040 |
| ILPP Distance | 0.042 | 0.009 | 4.830 | $1 \times 10^{-6}$ | 0.025 | 0.059 |
| Age | -0.051 | 0.01 | -5.090 | $4 \times 10^{-7}$ | -0.071 | -0.031 |
| | **Prediction of Corrected Temporal RNFL Thickness** | | | | | |
| Optic Nerve Tortuosity | 0.020 | 0.006 | 3.200 | 0.001 | 0.008 | 0.033 |
| ILPP Distance | 0.175 | 0.009 | 19.600 | $2 \times 10^{-85}$ | 0.158 | 0.193 |
| Age | -0.051 | 0.01 | -4.864 | $1 \times 10^{-6}$ | -0.071 | -0.030 |
| | **Prediction of Nasal RNFL Thickness** | | | | | |
| Optic Nerve Tortuosity | 0.019 | 0.007 | 2.732 | 0.006 | 0.005 | 0.032 |
| ILPP Distance | 0.136 | 0.009 | 15.459 | $7 \times 10^{-54}$ | 0.119 | 0.153 |
| Age | -0.092 | 0.01 | -9.109 | $8 \times 10^{-20}$ | -0.112 | -0.072 |
| | **Prediction of Corrected Nasal RNFL Thickness** | | | | | |
| Optic Nerve Tortuosity | 0.020 | 0.006 | 3.146 | 0.002 | 0.008 | 0.033 |
| ILPP Distance | 0.211 | 0.009 | 23.575 | $7 \times 10^{-123}$ | 0.193 | 0.228 |
| Age | -0.086 | 0.01 | -8.256 | $2 \times 10^{-16}$ | -0.106 | -0.065 |

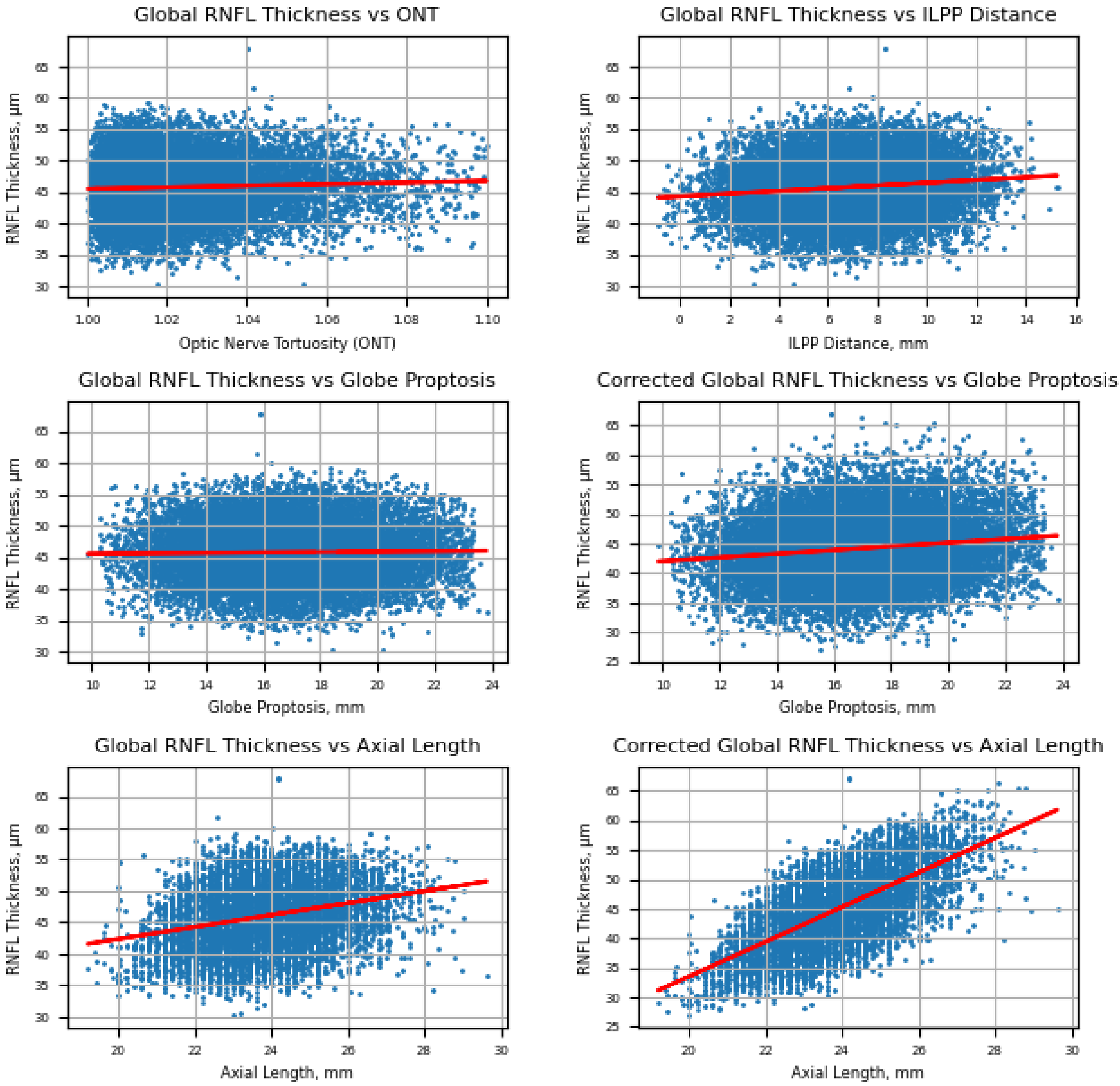

**Supplementary Figure 1.** Scatter plots of Retinal Nerve Fiber Layer (RNFL) thickness, with and without correction for ocular magnification, against features of the orbit in the general population

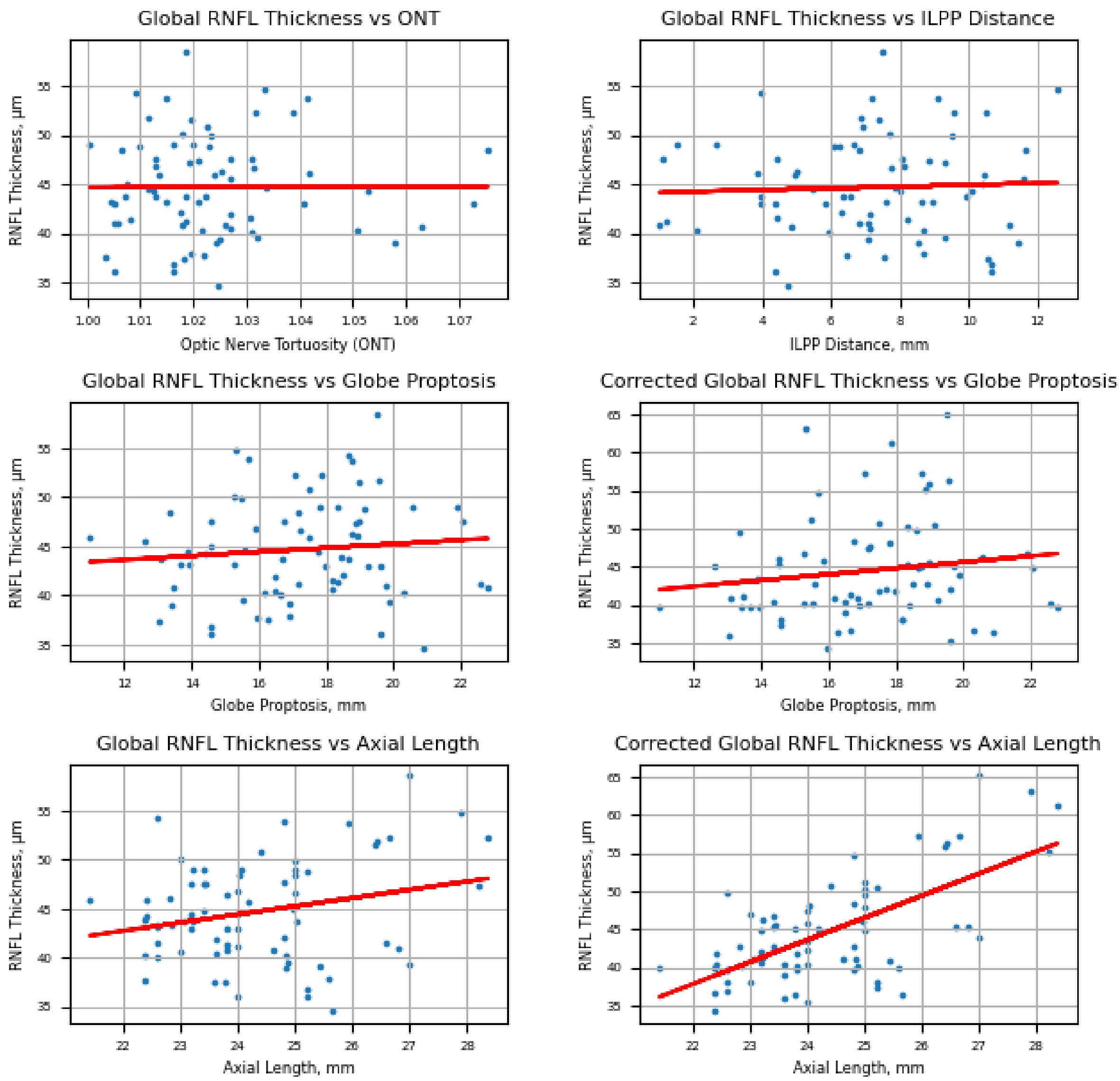

**Supplementary Figure 2.** Scatter plots of Retinal Nerve Fiber Layer (RNFL) thickness, with and without correction for ocular magnification, against features of the orbit in the primary open-angle glaucoma subpopulation

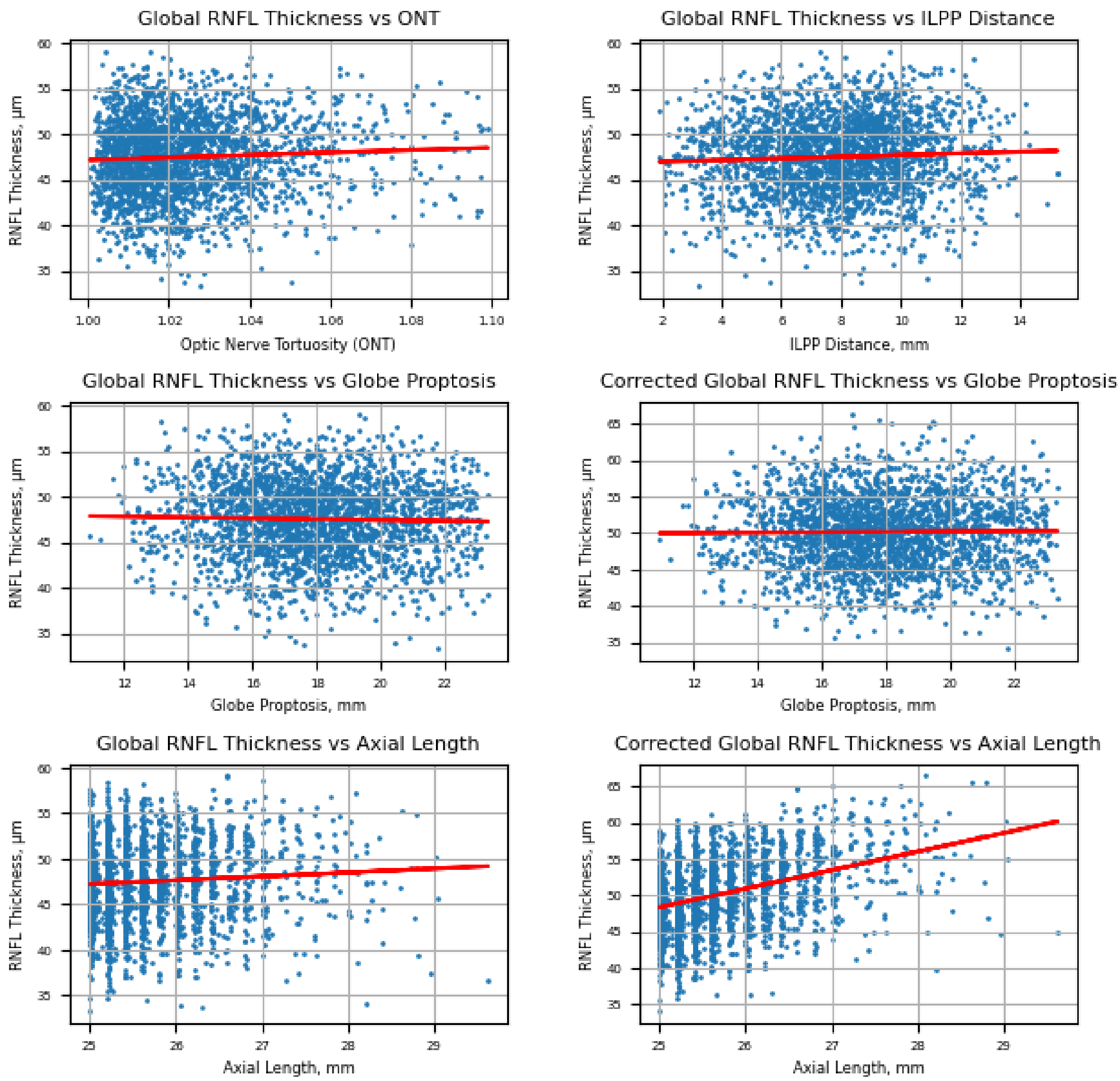

**Supplementary Figure 3.** Scatter plots of Retinal Nerve Fiber Layer (RNFL) thickness, with and without correction for ocular magnification, against features of the orbit in the myopic subpopulation

**Supplementary Note 1.**

In the UKBiobank, diagnostic information was recorded in the form of International Classification of Disease (ICD) codes. We used the following codes to identify populations of subjects with different types of glaucoma as detailed in Supplementary Table 8.

**Supplementary Table 8**. Glaucoma subpopulations

| ICD-10 Code | ICD Code Description | Number of Subjects | Number of Eyes |
|---|---|---|---|
| H400 | Glaucoma suspect | 40 | **73** |
| H401 | Primary open-angle glaucoma | 44 | 72 |
| H402 | Primary angle-closure glaucoma | 19 | 36 |
| H409 | Glaucoma, unspecified | 111 | 189 |

Due to different mechanisms underlying different glaucoma types, we only included subjects with primary open-angle glaucoma in the glaucoma subpopulation in the main text analysis. Differentiated analysis for other glaucoma types revealed broadly similar results when discounting statistically insignificant findings, as shown in the following Supplementary Tables 9 and 10.

**Supplementary Table 9**. Correlation of Corrected Global RNFL Thickness with various Orbital Features

| | **Primary Angle-Closure Glaucoma Subpopulation** | | **Unspecified Glaucoma Subpopulation** | | **Glaucoma Suspect Subpopulation** | |
|---|---|---|---|---|---|---|
| **Orbital Feature** | **Pearson r** | **p-value** | **Pearson r** | **p-value** | **Pearson r** | **p-value** |
| Optic Nerve Tortuosity | 0.038 | 0.83 | 0.152 | 0.04 | 0.191 | 0.11 |
| ILPP Distance | 0.306 | 0.07 | 0.273 | $1 \times 10^{-4}$ | 0.197 | 0.10 |
| Globe Proptosis | -0.133 | 0.44 | 0.022 | 0.76 | 0.245 | 0.04 |
| Axial Length | 0.391 | 0.02 | 0.470 | $9 \times 10^{-12}$ | 0.660 | $2 \times 10^{-10}$ |
| Age | 0.097 | 0.57 | -0.226 | 0.002 | 0.155 | 0.19 |

**Supplementary Table 10**. Linear GEE Regression Results to Predict Corrected Global RNFL Thickness

| | Coef. | Std Err | z | P>\|z\| | 95% CI | |
|---|---|---|---|---|---|---|
| | **Primary Angle-Closure Glaucoma Subpopulation** | | | | | |
| Optic Nerve Tortuosity | -0.036 | 0.081 | -0.451 | 0.652 | -0.194 | 0.122 |
| ILPP Distance | 0.286 | 0.146 | 1.963 | 0.050 | 0.001 | 0.572 |
| Age | 0.054 | 0.163 | 0.329 | 0.742 | -0.266 | 0.373 |
| | **Unspecified Glaucoma Subpopulation** | | | | | |
| Optic Nerve Tortuosity | 0.066 | 0.066 | 1.005 | 0.32 | -0.063 | 0.195 |
| ILPP Distance | 0.252 | 0.088 | 2.860 | 0.004 | 0.079 | 0.425 |
| Age | -0.216 | 0.078 | -2.783 | 0.005 | -0.369 | -0.064 |
| | **Glaucoma Suspect Subpopulation** | | | | | |
| Optic Nerve Tortuosity | -0.053 | 0.438 | -0.121 | 0.90 | -0.911 | 0.805 |
| ILPP Distance | 0.122 | 0.231 | 0.530 | 0.60 | -0.330 | 0.574 |
| Age | -0.004 | 0.071 | -0.052 | 0.96 | -0.142 | 0.135 |

**Supplementary Note 2.**

To check if eyes were in primary gaze, we measured the gaze angle, defined as the angle between the central orbital axis and the interzygomatic line. We manually examined the MRI scans of all subjects with eyes that had a gaze angle deviating from a right angle by more than five degrees and excluded subjects with eyes that deviated from primary gaze.

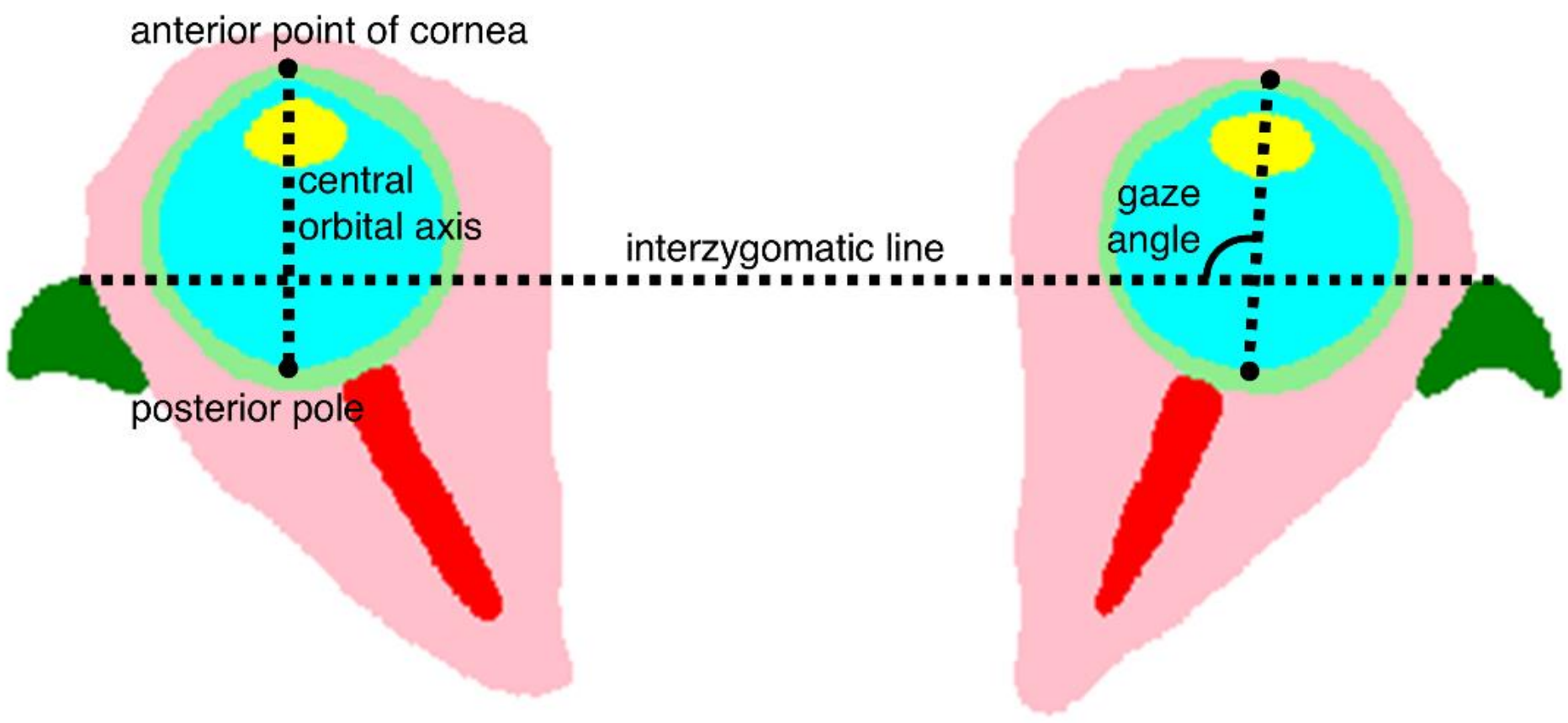

**Supplementary Figure 4.** Gaze Angle was defined as the angle between the central orbital axis and interzygomatic line. The gaze angle should be 90 degrees in subjects at primary gaze.

**Supplementary Note 3.**

Mutual information scores were used to evaluate the dependence between variables. Mutal information captures non-linear relationships and is thus, useful for detecting non-linear dependencies that might be missed by correlation-based methods. While the absolute values of mutual information scores are arbitrary, they can be used to compare the relative importance of features within a specific dataset by calculating mutual information between a target variable and each feature. A score of 0 indicates independence, while higher scores indicate stronger dependence.

The mutual information scores, as shown in Supplementary Table 8, indicated that Globe Proptosis and interzygomatic line-to-posterior pole (ILPP) Distance had similar amounts of mutual information with RNFL thickness, and more mutual information with RNFL thickness than other orbit features (Supplementary Table 11). Nerve tortuosity followed closely in the ranking of features sharing most mutual information with RNFL thicknesses.

**Supplementary Table 11**. Mutual Information Scores of Orbit Features with Global RNFL Thickness

| **Orbital Feature** | **General Population** | | **Glaucoma Subpopulation *** | | **Myopic Subpopulation** | |
|---|---|---|---|---|---|---|
| Globe Proptosis | 2.87 | 1 | 4.06 | 2 | 4.93 | 1 |
| ILPP Distance | 2.74 | 2 | 4.12 | 1 | 4.89 | 2 |
| Nerve Tortuosity | 2.63 | 3 | 4.06 | 3 | 4.70 | 3 |
| Axial Length | 1.50 | 4 | 3.85 | 4 | 3.44 | 4 |
| Age | 0.86 | 5 | 3.03 | 5 | 2.65 | 5 |

**Glaucoma Subpopulation comprises subjects with primary open-angle glaucoma*

As Globe Proptosis and ILPP Distance were highly collinear, only one could be included 6as a predictor in the GEE model. Given that Globe Proptosis and ILPP Distance both had similar amounts of mutual information with RNFL thickness, but ILPP Distance was significantly more strongly correlated with RNFL thickness than Globe Proptosis (Tables 3 and 4), i.e. stronger linear relationship, thus, ILPP distance was included as a predictor in the linear GEE model rather than Globe Proptosis.